\documentclass[11pt]{article}

\usepackage{amsmath,amssymb,graphicx,slashbox}

\def\mydate{March 31, 2010}
\def\ignore#1{{}}

\newcommand{\beeq}{\begin{equation}}
\newcommand{\eneq}{\end{equation}}
\newcommand{\beqn}{\begin{eqnarray}}
\newcommand{\eeqn}{\end{eqnarray}}

\def\la{\raise.16ex\hbox{$\langle$}\lower.16ex\hbox{}  }
\def\ra{\raise.16ex\hbox{$\rangle$}\lower.16ex\hbox{} }
\def\go{\rightarrow}

\def\onehalf{ \hbox{${1\over 2}$} }

\def\eff{{\rm eff}}

\def\EM{{\rm EM}}

\def\KK{{\rm KK}}

\def\hatH{\hat\theta_H}

\def\psibar{ \psi \kern-.65em\raise.6em\hbox{$-$} }
\def\psibarl{ \psi \kern-.65em\raise.6em\hbox{$-$} \lower.6em\hbox{} }

\begin{document}

{\small \noindent \mydate    \hfill OU-HET 660/2010}

\vskip 1.cm 

\begin{center}

\baselineskip=20pt plus 1pt minus 1pt

{\LARGE \bf Stable Higgs Bosons\\
\vskip 5pt
-- new candidate for cold dark matter\footnote{To appear in the Proceedings of
{\it ``The 10th. International Symposium on 
Origin of Matter and Evolution of the Galaxies''} (OMEG10),
March 8-10, 2010, RCNP, Osaka, Japan. } --}

\vskip .3cm

{\large Yutaka Hosotani}

{\small \it Department of Physics, Osaka University, Toyonaka, Osaka 560-0043, Japan}

\end{center}

\begin{abstract}
{\small
The Higgs boson is in the backbone of the standard model of electroweak interactions.  
It must exist in some form for achieving unification of interactions.
In the gauge-Higgs unification scenario the  Higgs boson
becomes a part of the extra-dimensional component of  gauge fields.
The Higgs boson becomes absolutely
stable in a class of the gauge-Higgs unification models, 
serving  as a promising candidate for cold dark matter in the universe.  
The observed relic abundance of  cold dark matter
is obtained with the Higgs mass  around 70$\,$GeV.  
The Higgs-nucleon scattering cross section is found to be
close to the recent CDMS II and XENON10 bounds 
in the direct detection of dark matter.
In  collider experiments stable Higgs bosons are produced
in a pair,   appearing as missing energies and momenta so that
the way of detecting Higgs bosons must be altered.}
\end{abstract}

\baselineskip=11pt plus 1pt minus 1pt
{\small {\bf Keywords:} 
dark matter, Higgs boson, gauge-Higgs unification, Hosotani mechanism}


\vskip 15pt

\baselineskip=13pt plus 1pt minus 1pt

\section{Introduction}

What constitutes the dark matter in the universe?
Will the Higgs bosons be found at Tevatron and LHC as described in the 
standard model of electroweak interactions?
These are two of the most urgent and important questions in current  physics.
The dark matter and the Higgs boson,  we do not know what they really are.
In this talk I would like to present a new scenario that these two mystery 
particles are really the same.\cite{YHscgt09, YHsusy08} 

In the gauge-Higgs unification scenario the 4D Higgs boson is
identified with  a part of the gauge field  in an extra dimension of 
spacetime.\cite{YH1}-\cite{Lim1}
When the extra dimensional space is not simply connected, the 4D
Higgs boson becomes absolutely stable under a certain conditions.
As a consequence Higgs bosons become dark matter of the universe.
The relic abundance of the Higgs bosons in the universe is estimated
with the Higgs boson mass $m_H$ being an only relevant free parameter.  
The WMAP data for the mass density of cold dark matter  is reproduced
with $m_H \sim 70\,$GeV.\cite{HKT}  The light Higgs boson mass does not 
contradict with the LEP2 bound $m_H > 114\,$GeV.  In the gauge-Higgs 
unification scenario the $ZZH$ coupling vanishes so that the process
$e^+ e^- \go Z \go ZH$ does not take place, the LEP2 bound thus being evaded.

\section{Why do we need ``Higgs''?}

The standard model of strong and electroweak interactions is extremely 
successful.  The framework of gauge interactions, in both strong 
and electroweak interactions, has been firmly established to high accuracy.
In the electroweak sector, however, there must be an additional particle
yet to be discovered.  It is the Higgs boson.   Without it
the electroweak unification cannot be achieved in the standard model.

Why do we need the Higgs boson?  In the unification scenario 
the theory has larger symmetry than the observed world.  The electroweak
unified theory has $SU(2)_L \times U(1)_Y$ gauge symmetry,
whereas the only $U(1)_\EM$ gauge symmetry, the Maxwell electromagnetic
gauge invariance, is observed at low energies.  The electroweak 
$SU(2)_L \times U(1)_Y$ gauge symmetry must be spontaneously broken to
the electromagnetic $U(1)_\EM$ gauge symmetry.  In the standard model
this symmetry breaking is induced by dynamics of the Higgs boson $\phi$.

The Higgs boson develops a vacuum expectation value $\la \phi (x)  \ra = v$
as a result of the spontaneous symmetry breaking.
Its fluctuation $H(x)$ in $\phi(x) = v + H(x)$ is the Higgs boson to be discovered.
The Higgs field couples to quarks and leptons with $y (v+H) \bar \psibar  \psi$ where
$y$ stands for a Yukawa coupling.   Non-vanishing $v$ generates a fermion 
mass $m = yv$.  It implies that the Higgs boson decays into a fermion-antifermion 
pair with a coupling $y=m/v$.  The Higgs boson becomes necessarily unstable 
in the standard model.

The Higgs boson has not been found yet.  We have  measured 
neither the Yukawa couplings nor the Higgs couplings to $W$ and $Z$.
We do not have direct information on the Higgs sector.

Unlike the gauge sector the Higgs sector in the standard model does not
have a principle governing the interactions.  All the couplings of the Higgs
boson are arbitrary.  They are determined to fit the experimental data.
Further it is known that the Higgs boson mass is unstable against radiative
corrections.   These are unsatisfactory features in the standard model.

\section{Gauge-Higgs unification}

The gauge-Higgs unification scenario gives a solution to those problems.
The 4D Higgs field is unified with 4D gauge fields in  gauge
theory in higher dimensions a la Kaluza-Klein.  The gauge principle governs
the Higgs interactions.

The four-dimensional components $A_\mu$ of the gauge potential $A_M$
contain 4D gauge fields, whereas the extra-dimensional component $A_y$
contains the 4D Higgs field.
\beqn
&A_M(x,y) ~ \hbox{in higher dimensions}& \cr
\noalign{\kern 5pt}
&\overbrace{~ \hskip 5.5cm ~}&\cr
&\hbox{four-dim.\ components}~A_\mu \hskip 1.2cm 
       \hbox{extra-dim.\ component}~A_y~& \cr
&\bigcup \hskip 5cm \bigcup&\cr
&\hbox{4D gauge fields}~(\gamma , W, Z) \hskip 2cm  
\hbox{4D Higgs field}~(H) ~~~~~&   \label{scheme1} \\ 
&\hskip 5.2cm \hbox{as an Aharonov-Bohm phase}& \cr
&\hskip 5.4cm \Downarrow& \cr
&\hskip 5.4cm \hbox{EW symmetry breaking}&
\nonumber
\eeqn

The scenario works well when the extra-dimensional space is 
non-simply-connected.\cite{YH1}
In that case there appears an Aharonov-Bohm (AB) phase,
which is given by the phase of the path-ordered integral along a non-contractible 
loop $C$  in the extra dimension 
\beeq
P \exp \Big\{ ig \int_C dy \, A_y (x, y) \Big\} ~.
\eneq
The AB phase is a physical degree of freedom remaining  in configurations
with vanishing field strengths $F_{MN}=0$.
The 4D neutral Higgs field is nothing but  4D  fluctuation of the relevant AB phase
$\theta_H$.
At the classical level the value of $\theta_H$ is undetermined as $F_{MN}=0$
so that the Higgs field is massless.
Quantum corrections make the effective potential $V_\eff (\theta_H)$ 
nontrivial.  When $V_\eff$ is minimized at a nontrivial $\theta_H \not= 0$, 
the gauge symmetry is  spontaneously broken.
At the same time a finite Higgs mass $m_H$ is dynamically generated. 
This should be contrasted to the earlier attempt of gauge-Higgs unification
on $M^4 \times S^2$, in which all $W$, $Z$, and Higgs bosons  have
masses of the Kaluza-Klein mass scale $m_\KK$.\cite{Fairlie1}-\cite{YH4}

A realistic, promising model is constructed with $SO(5) \times U(1)$ gauge symmetry
in the five-dimensional Randall-Sundrum (RS) warped spacetime.\cite{Agashe1}-\cite{HNU}
The metric of the RS spacetime \cite{RS1} is given by 
\beeq
ds^2   = e^{-2\sigma(y)}\eta_{\mu\nu} dx^\mu dx^\nu + dy^2 , \qquad
\label{metric1}
\eneq
where $\eta_{\mu\nu} =\textrm{diag}(-1,1,1,1)$,
$\sigma(y)=\sigma(y+2L)$, and
$\sigma(y)=k|y|$ for $|y|\leq L$.
The fundamental region in the fifth dimension
is given by $0\leq y\leq L$.
The Planck  and TeV branes are located at $y=0$ 
and $y=L$, respectively.  
The bulk region $0 < y < L$ is  anti-de Sitter spacetime with a 
cosmological constant  $\Lambda = - 6k^2$.
The metric is specified with two parameters, the AdS curvature scale $k$ and 
the warp factor $w_L = e^{kL} \gg 1$.

The RS spacetime has the same topology as the orbifold $M^4 \times (S^1/Z_2)$. 
The spacetime points $(x^\mu, y)$, $(x^\mu, y+2L)$, and $(x^\mu, -y)$
are identified with each other.  Gauge potentials $A_M(x,y)$ are defined in 
covering space  $-\infty < y < \infty$.  
$A_M(x, -y)$ need not be the same  as  $A_M(x,y)$ in gauge theory .  
It may be twisted by a gauge transformation.  Only physical quantities
must be single-valued at $y$, $y+2L$, and $-y$.  This brings orbifold boundary conditions
at the Planck  and TeV  branes, which, in the model under consideration,  
break $SO(5) \times U(1)$ to $SO(4) \times U(1) \simeq SU(2)_L \times SU(2)_R \times U(1)$.  

Fermions in the bulk (between the two branes) belong to the vector
representation of $SO(5)$.  Two vector multiplets are necessary to describe,
say, top and bottom quarks.  In addition, brane fermions are introduced on the 
Planck brane,  in the $(\onehalf, 0)$ representation of 
$SU(2)_L \times SU(2)_R$.  A brane scalar on the Planck brane,
in the  $(0, \onehalf)$ representation, spontaneously breaks $SU(2)_R \times U(1)$
to $U(1)_Y$,   simultaneously generating mass couplings between
the bulk and brane fermions.  The surviving symmetry  is the 
standard electroweak (EW) symmetry $SU(2)_L \times U(1)_Y$.

There appear four AB phases, three of which are absorbed by $W$ and $Z$ bosons.
The remaining AB phase $\theta_H$ corresponds to the physical 4D neutral Higgs 
boson $H(x)$.   It appears in the combination
\beeq
\hatH (x) = \theta_H + \frac{H(x)}{f_H}
~~,~~
f_H = \frac{2}{g_A} \sqrt{\frac{k}{z_L^2 -1}}
\sim \frac{2}{\sqrt{kL}} \frac{m_\KK}{\pi g} ~~. 
\label{thetahat}
\eneq
Here the  Kaluza-Klein (KK) mass scale  is  
$m_\KK = \pi k z_L^{-1}$, and
the $SO(5)$ gauge coupling $g_A$ is related to the 4D $SU(2)_L$ weak 
coupling $g$ by $g=g_A/\sqrt{L}$.  
The theory is invariant under $\theta_H \go \theta_H + 2\pi$ as a result of 
the large gauge invariance.

\begin{figure}[tb]
\begin{center}
\includegraphics[height=4.5cm]{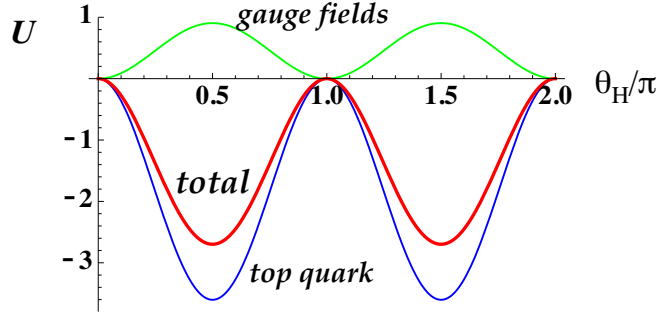}
\caption{Effective potential $V_\eff (\theta_H) = m_\KK^4 /(16 \pi^4) U (\theta_H)$  
with $z_L = 10^{10}$.  At its gobal minimum $\theta_H = \onehalf\pi$, 
the EW symmetry is spontaneously broken to $U(1)_\EM$.}
\label{potential}
\end{center}
\end{figure}

The value of $\theta_H$ is determined by the location of the global 
minimum of the effective potential $V_\eff (\theta_H)$.
$V_\eff (\theta_H)$ at the one loop level  is depicted in fig.\ \ref{potential}.
It has global minima at $\theta_H = \pm \onehalf \pi$.  The contribution from top
quarks dominates over those from gauge fields.  Contributions from other 
quarks and leptons are negligible.   At $\theta_H = \pm \onehalf \pi$ 
the EW symmetry ($SU(2)_L \times U(1)_Y$) breaks down to $U(1)_\EM$.

\section{Effective theory}

In the effective theory of  low energy fields the Higgs field $H(x)$ enters 
always in the combination of $\hatH (x)$ in (\ref{thetahat}).   
The effective Lagrangian must be invariant under $\hatH (x) \go \hatH (x) + 2\pi$.
The Higgs interactions with the $W$, $Z$ bosons,  quarks and leptons  
at low energies are summarized as \cite{HKT, HK, Sakamura1}
\beeq
{\cal L}_\eff  = - V_\eff (\hatH) 
                 - m_W^2(\hatH) W^\dagger_\mu W^\mu
                 - \onehalf m_Z^2(\hatH) Z_\mu Z^\mu 
                 - \sum_{a,b} m^F_{ab}(\hatH) \psibar_a  \psi_b ~. 
\label{effective1}
\eneq
The effective potential $V_\eff (\hatH) $ arises 
at the one loop level.  The mass functions $m_{W,Z} (\hatH)$ and 
$ m^F_{ab}(\hatH)$, on the other hand, appear  at the tree level.

In the $SO(5) \times U(1)$ model under consideration these mass
functions are found, to good accuracy in the warped space,   to be
\cite{HK, SH1, HS2}
\beqn
&&\hskip -1cm 
m_W(\hatH) \sim \cos \theta_W m_Z(\hatH)  \sim
\frac{1}{2} g f_H \sin \hatH ~~, 
\quad
\bigg[ ~ \frac{1}{2}  g (v + H) ~~  \hbox{in SM}  \bigg]  ~,  \cr
\noalign{\kern 10pt}
&&\hskip -1cm
m_{ab}^F(\hatH)  \sim
y_{ab}^F  f_H \sin \hatH ~~, 
\quad 
\Big[ ~ y_{ab}^F  (v + H) ~~  \hbox{in SM} \Big]   ~.
\label{effective2}
\eeqn
We have listed the formulas in the standard model in brackets.
It is seen that $v+H$ in the standard model is replaced approximately
by $f_H \sin \hatH$  in the gauge-Higgs unification.  

The masses of the $W$, $Z$ bosons and fermions are given by
$m_W = \onehalf g f_H \sin \theta_H$, 
$m_Z= m_W/\cos \theta_W$,  and 
$m_{ab}^F = y_{ab}^F f_H \sin \theta_H$.  
The Higgs couplings to $W$, $Z$ and 
fermions deviate from those in the standard model.  
It follows from (\ref{effective2}) that 
\beqn
\begin{pmatrix} WWH \cr ZZH \cr \hbox{Yukawa} \end{pmatrix}
&=& \hbox{SM} \times \cos \theta_H ~~, \cr
\noalign{\kern 5pt}
\begin{pmatrix} WWHH \cr ZZHH  \end{pmatrix}
&=& \hbox{SM} \times \cos 2 \theta_H ~~.
\label{coupling2}
\eeqn
As shown in the previous section $\theta_H= \onehalf \pi$ is dynamically chosen
so that $f_H \sim 246\,$GeV. 
All of the $WWH$, $ZZH$ and Yukawa couplings vanish at $\theta_H= \onehalf \pi$,
implying that the Higgs boson cannot decay.
This is an exact result as shown below.  

\section{Absolutely Stable Higgs bosons}

One of the astonishing results from the gauge-Higgs unification scenario is that Higgs 
bosons become absolutely stable.  It follows from  symmetry and dynamics
of the theory.

First there exists the mirror reflection symmetry in the fifth dimension;
$(x^\mu, y) \go (x^\mu, -y)$.
It implies that physics is invariant under
\beeq
\hatH (x) = \theta_H + \frac{H(x)}{f_H} \go  - \hatH(x)
\label{reflect2}
\eneq
while all other SM particles remain unchanged.

Secondly there arises  enhanced gauge invariance.  In our model the bulk fermions
are all in the vector representation of $SO(5)$.  There is no matter field in the spinor 
representation of $SO(5)$.  Brane fermions are introduced at the Planck brane only.
In this case  the theory turns out invariant under 
\beeq
\theta_H \go \theta_H + \pi ~.
\label{shift2}
\eneq
All physical quantities become periodic in $\theta_H$ with a reduced period $\pi$.

Now recall that the dynamics select $\theta_H = \onehalf \pi$.  Then we have equivalence
relations
\beeq
\frac{\pi}{2} + \frac{H}{f_H} ~ \Leftrightarrow~
- \frac{\pi}{2} -  \frac{H}{f_H}  ~\Leftrightarrow~
\frac{\pi}{2} - \frac{H}{f_H} 
\label{equivalence}
\eneq
where the first and second equivalece relations follow from (\ref{reflect2}) 
and (\ref{shift2}), respectively.  The net result is the change in the sign of the 
4D Higgs field $H(x)$.  All other SM fields remain unchanged.   
There emerges  the $H$-parity invariance.   
\beqn
\hbox{Higgs boson~}   : 
&&   \hbox{H-parity}~~ - \cr
\noalign{\kern 5pt}
\hbox{all other SM particles} ~:
&&   \hbox{H-parity}~ + .
\label{Hparity}
\eeqn
Among low energy fields only the Higgs field is $H$-parity odd.
The Higgs boson becomes absolutely stable, protected by the $H$-parity conservation.

\section{Higgs dark matter}

As an important consequence,   abosolutely stable Higgs bosons become 
cold dark matter (CDM)  in the present universe.  
They are copiously produced in the very early universe.  
As the annihilation rates of Higgs bosons fall below the expansion rate 
of the universe,  the annihilation processes get effectively  frozen
and the remnant Higgs bosons become dark matter.\cite{HKT} 

In the gauge-Higgs unification scenario 
the annihilation rates can be estimated from the effective Lagrangian
(\ref{effective1}) with (\ref{effective2}), once the Higgs boson mass   
$m_H$ is given.\cite{DM3, BHS}  
If $m_H > m_W$, the dominant annihilation modes are  $HH \go WW, \, ZZ$.
The rate is large so that the resultant relic abundance becomes very small.
For $m_H < m_W$ the relevant annihilation modes are 
$HH \go W^* W^*, \, Z^* Z^*$, $b \bar b, c \bar c, \tau \bar \tau$, and
$ gg$.  Here $W^*$ ($Z^*$) denotes virtual $W$ ($Z$) which subsequently
decays into a fermion pair.  Annihilation into a gluon ($g$) pair takes place 
through a top quark  loop.

\begin{figure}[bt]
\begin{center}
\includegraphics[height=5.5cm]{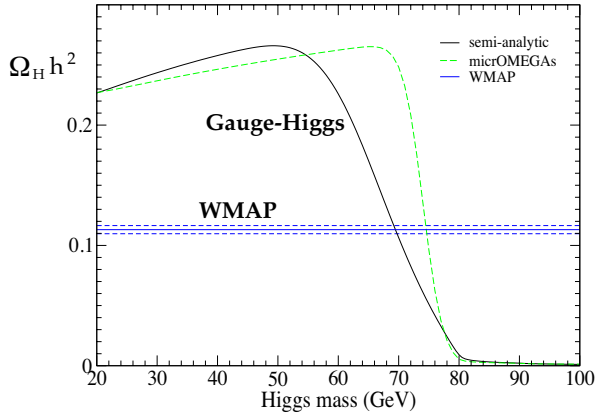}
\caption{Relic abundance of the Higgs dark matter.
The solid curve is obtained by the semi-analytic formulas.
The horizontal band is the WMAP data 
$\Omega_{\rm CDM} h^2 = 0.1131 \pm 0.0034$.}
\label{relic1}
\end{center}
\end{figure}

The relic abundance of  Higgs bosons  as a function of 
$m_H$  is depicted in fig.\  \ref{relic1}.
$\Omega_H h^2$ determined from the WMAP data
is reproduced  with $m_H \sim 70\,$GeV.
With $m_H = 70\,$GeV, the freeze-out temperature is $T_f \sim 3\,$GeV.
The relative contributions of the $HH \go W^* W^*$ and $b \bar b$ modes
are 61\% and 34\%, respectively.

The Higgs dark matter can be detected by observing Higgs-nucleon elastic 
scattering  $HN \go HN$.
The  spin-independent (SI) Higgs-nucleon scattering cross section is found to be
\beeq
\sigma_{\rm SI} \simeq
\frac{1}{4\pi} \bigg( \frac{2 + 7 f_N}{9} \bigg)^2 
\frac{m_N^4}{f_H^4 ( m_H + m_N)^2} ~,
\label{SIrate}
\eneq
where $f_N=\sum_{q=u,d,s}  \langle N| m_q\bar q q | N \rangle / m_N$
is estimated in a range $ 0.1 \sim 0.3$.
With $m_H = 70\,$GeV and $f_H=246\,$GeV, $\sigma_{\rm SI}$ is
estimated to be 
$\sigma_\mathrm{SI} \simeq (1.2 \sim 2.7) \times 10^{-43}\,{\rm cm}^2$.

The present experimental upper bounds for
the spin-independent WIMP-nucleon cross sections come from 
CDMS II \cite{CDMS} and XENON10 \cite{XENON}.
From the recent CDMS II data
$\sigma_\mathrm{SI} \lesssim 7 \times 10^{-44}\,\mathrm{cm}^2$ 
at 90 \% CL with the WIMP mass $70\,$GeV.
With many uncertainties and ambiguity in the analysis taken into account, 
this does not necessarily mean that the present  gauge-Higgs unification model
is excluded.   The next generation experiments for the direct detection of 
WIMP-nucleon scattering are awaited to pin down the rate.

The  scenario under discussion,  in which the Higgs bosons responsible for the 
EW symmetry breaking become the dark matter, differs from 
that of refs.\ \cite{Barbieri, Gustafsson} where an additional Higgs 
doublet with odd parity is introduced by hand.
It also differs from the Kaluza-Klein (KK) dark matter scenario in which additional fields 
with odd KK parity become dark matter.\cite{KKdark1}-\cite{KKdark5}
Unlike those models the relic abundance and the direct detection rate are
unambiguously estimated in the present model once $m_H$ is known.

\section{At colliders}

Higgs bosons are produced in pairs at colliders.  Typical production processes are
$Z^* \go ZHH$, $W^* \go WHH$, $WW \go HH$, $ZZ \go HH$, and $gg \go HH$.
Higgs bosons are stable so that they appear as missing energies and momenta.
The appearance of two particles of missing energies 
and momenta in the final state makes experiments hard, but not impossible.  
These events must be distinguished from those involving neutrinos.
In all of the experiments performed so far, Higgs bosons were
searched by trying to  identify their decay products.  Since Higgs bosons are stable,
this way of searching Higgs bosons must be altered.

There arise deviations in the gauge couplings of quarks and leptons, 
resulting in violation of the universality.\cite{HNU}  
The deviations in the $W$ and $Z$ couplings are tiny (0.1\% to 1\%)
except for top quarks.
An important prediction is given for the forward-backward asymmetry
in the $e^+ e^-$ collisions on the $Z$ pole.\cite{uekusa}
CP violation,   anomalous magnetic moment, 
and  electric dipole moment in gauge-Higgs unification have been also discussed.\cite{Lim4}

\section{Summary}

In the gauge-Higgs unification scenario the  Higgs boson is unified with
gauge fields.  The Higgs bosons naturally become stable in the 
$SO(5) \times U(1)$ unification. They become the cold dark matter of the 
present universe.    The mass of the Higgs boson  is determined to be 
$\sim 70\,$GeV from the WMAP data.  The direct detection rate for the 
WIMP(Higgs)-nucleon scattering is predicted to be very close to the current limit.

In collider experiments Higgs bosons appear as missing energies and momenta
so that the way of searching Higgs bosons must be altered.    The gauge-Higgs 
unification scenario also leads to slight deviations  from the standard model
in the electroweak gauge couplings, which can be checked in  experiments.

To summarize, the gauge-Higgs unification scenario is promising. 
The stable Higgs boson is a new candidate for cold dark matter.

\vskip 20pt

\leftline{\large \bf Acknowledgment}
\vskip 10pt

This work was supported in part 
by  Scientific Grants from the Ministry of Education and Science, 
Grant No.\ 20244028, Grant No.\ 20025004,  and Grant No.\ 50324744.

\newpage



\renewenvironment{thebibliography}[1]
         {\begin{list}{[$\,$\arabic{enumi}$\,$]}  
         {\usecounter{enumi}\setlength{\parsep}{0pt}
          \setlength{\itemsep}{0pt}  \renewcommand{\baselinestretch}{0.95}
          \settowidth
         {\labelwidth}{#1 ~ ~}\sloppy}}{\end{list}}

\end{document}